\documentclass[sigconf,natbib=true,anonymous=false]{acmart}

\usepackage{amsmath}
\usepackage{multirow} 
\usepackage{multicol} 
\usepackage{arydshln}
\usepackage{enumitem}
\usepackage{amsthm,amsmath}
\usepackage{mathrsfs}
\usepackage{balance}
\AtBeginDocument{%
  \providecommand\BibTeX{{%
    \normalfont B\kern-0.5em{\scshape i\kern-0.25em b}\kern-0.8em\TeX}}}

\copyrightyear{2022}
\acmYear{2022}
\setcopyright{acmcopyright}
\acmConference[CIKM '22]{Proceedings of the 31st ACM
International Conference on Information and Knowledge Management}{October
17--21, 2022}{Atlanta, GA, USA}
\acmBooktitle{Proceedings of the 31st ACM International Conference on Information
and Knowledge Management (CIKM '22), October 17--21, 2022, Atlanta, GA, USA}
\acmPrice{15.00}
\acmDOI{10.1145/3511808.3557674}
\acmISBN{978-1-4503-9236-5/22/10}

\acmSubmissionID{sp1040}

\begin{document}

\title{Prototypical Contrastive Learning and Adaptive Interest Selection for Candidate Generation in Recommendations}

% \author{Ningning Li, Qunwei Li, Xichen Ding, Shaohu Chen and Wenliang Zhong}
% \email{nelson.lnn, qunwei.qw, xichen.dxc, shaohu.csh, yice.zwl@antgroup.com }

\author{Ningning Li}
\orcid{1234-5678-9012}
\affiliation{%
  \institution{Ant Group}
  \city{Hangzhou}
%   \state{Zhejiang}
  \country{China}
}
\email{nelson.lnn@antgroup.com}

\author{Qunwei Li}
\orcid{1234-5678-9012}
\affiliation{%
  \institution{Ant Group}
  \city{Hangzhou}
%   \state{Zhejiang}
  \country{China}
}
\email{qunwei.qw@antgroup.com}

\author{Xichen Ding}
\orcid{1234-5678-9012}
\affiliation{%
  \institution{Ant Group}
  \city{Beijing}
  \country{China}
}
\email{xichen.dxc@antgroup.com}

\author{Shaohu Chen}
\orcid{1234-5678-9012}
\affiliation{%
  \institution{Ant Group}
  \city{Beijing}
  \country{China}
}
\email{shaohu.csh@antgroup.com}

\author{Wenliang Zhong}
\orcid{1234-5678-9012}
\affiliation{%
  \institution{Ant Group}
  \city{Hangzhou}
%   \state{Zhejiang}
  \country{China}
}
\email{yice.zwl@antgroup.com}

\renewcommand{\shortauthors}{Ningning Li et al.}

\begin{abstract}

Deep Candidate Generation plays an important role in large-scale recommender systems. It takes user history behaviors as inputs and learns user and item latent embeddings for candidate generation. In the literature, conventional methods suffer from two problems. First, a user has multiple embeddings to reflect various interests, and such number is fixed. However, taking into account different levels of user activeness, a fixed number of interest embeddings is sub-optimal. For example, for less active users, they may need fewer embeddings to represent their interests compared to active users. Second, the negative samples are often generated by strategies with unobserved supervision, and similar items could have different labels. Such a problem is termed as class collision. In this paper, we aim to advance the typical two-tower DNN candidate generation model. Specifically, an Adaptive Interest Selection Layer is designed to learn the number of user embeddings adaptively in an end-to-end way, according to the level of their activeness. Furthermore, we propose a Prototypical Contrastive Learning Module to tackle the class collision problem introduced by negative sampling. Extensive experimental evaluations show that the proposed scheme remarkably outperforms competitive baselines on multiple benchmarks.

\end{abstract}



%% Self-supervised Learning, 

\maketitle

\section{Introduction}

\begin{figure}[h]
  \centering
  \includegraphics[width=8.4cm]{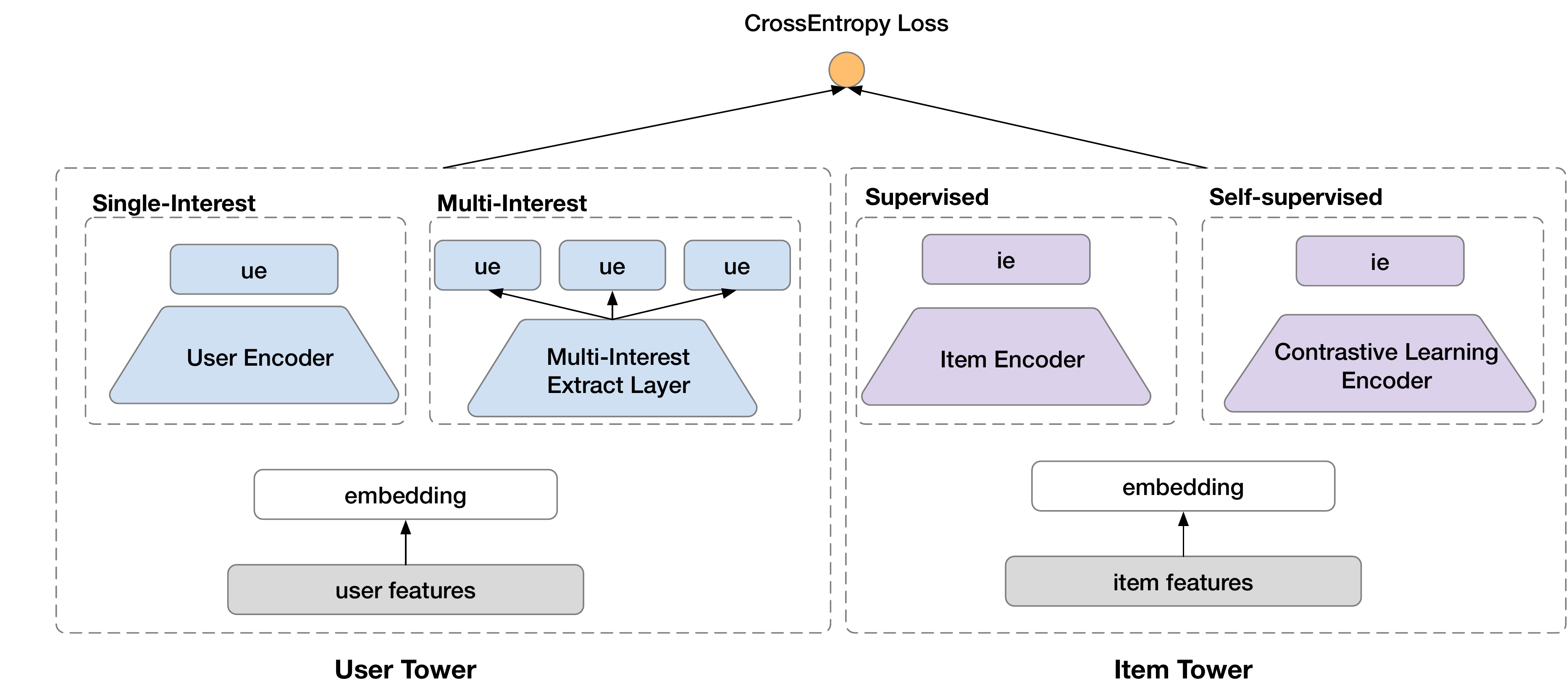}
  \caption{Deep Candidate Generation Model}
  \Description{}
  \label{fig:dcg}
\end{figure}

% 背景介绍
% In practice, recommender systems usually consist of two stages, candidate generation stage and ranking stage. The candidate generation stage is designed to retrieve top-$N$ candidate items from all available items, while the ranking stage aims at giving a more precise ranking over the candidates. Deep candidate generation (DCG) models have been widely adopted for many large web-scale recommender systems \cite{YouTubeDNN,FacebookNegativeSample,MIND,FuyuLv2019SDMSD,TianshengYao2020SelfsupervisedLF}.
Deep candidate generation (DCG) models have been widely adopted for many large web-scale recommender systems and is designed to retrieve top-$N$ candidate items from all available items \cite{YouTubeDNN,FacebookNegativeSample,MIND,FuyuLv2019SDMSD,TianshengYao2020SelfsupervisedLF}.
% \cite{YouTubeDNN,FacebookNegativeSample,MIND,FuyuLv2019SDMSD,TianshengYao2020SelfsupervisedLF,QijieShen2022DeepIH}.
A DCG model consists of two towers: the user tower and the item tower, to model the interactions between a user and an item, which is shown in Figure \ref{fig:dcg}. 
%It takes user and item features as inputs and learns user and item embeddings through neural networks with independent parameters. 
The user and item are represented as vectors in a common embedding space and the task of DCG is a classification problem with the cross-entropy loss to predict whether there is an interaction between the user and the item \cite{WalidKrichene2018EfficientTO,XinyangYi2019SamplingbiascorrectedNM}.
In such an architecture, the distances between the user and the candidate items can be calculated with such embeddings, and finding the nearest $N$ items to the user is called candidate generation in recommender systems. The retrieval of such $N$ nearest items from a large-scale corpus of items for a user is followed by ranking only these $N$ items. 
%Such a framework is efficient in deployment as it reduces the number of candidates to potential items of interest to a user for the time-consuming process of ranking. 
Additionally, candidate generation with efficient indexing algorithms such as LSH is sublinear in complexity \cite{JDH17faiss,MIPS}.

% 相关工作以及缺点
There are some existing works trying to learn efficient and effective %user and item 
embeddings. 
For user tower, there are mainly two methods as is shown in Figure \ref{fig:dcg}.
% (1)单向量 (2)多向量
(a) Single-Interest methods encode user raw features to a single dense vector through neural networks to represent user's interest \cite{YouTubeDNN,FuyuLv2019SDMSD}, which is the most common method.
(b) Multi-Interest methods leverage capsule networks or sequential recommendation networks, which are also called multi-interest extract layer, to extract multiple embeddings from user's historical behaviors to represent their diverse interests \cite{MIND,YukuoCen2020ControllableMF}.

% 问题：用户兴趣是fix住的
However, these DCG models ignore user's difference in activeness and fix the number of user embeddings.
We argue that more active users should have more interest footprints compared to less active users and new users, and should have a larger number of user embeddings.
% (3) 使用SSL的工作
For item tower %in Figure \ref{fig:dcg}
, (a) supervised methods train the item encoder using observed supervision; (b) Self-supervised methods mainly uses contrastive learning to regularize some latent parameters, such that similar items are closer in the embedding space than dissimilar ones \cite{sslreg01,sslreg02,XinyangYi2019SamplingbiascorrectedNM,ContrastiveDebias,TianshengYao2020SelfsupervisedLF}. Thus, we can learn item embeddings respecting their similarity features.
% 问题: low-level + class collision

An item can have structural features. For example, the item ID is the low-level information and the category or the brand of items counts as high-level feature.
Unfortunately, typical contrastive learning methods has been shown that it is only capable of encoding low-level information of items rather than high-level semantic features. In DCG models, things can be even uglier because a tremendously large number of negative samples are generated by strategies with unobserved supervision and manually assigned labels, and many similar item pairs could have different labels and are undesirably separated apart in the embedding space, which is called the class collision problem \cite{JunnanLi2020PrototypicalCL, DataCollision}.

% 阐述我们的方法
In this paper, we propose a novel and effective framework to improve DCG model's performance.
% in recommender systems.
First, we add an Adaptive Interest Selection Layer (AISL) on top of the multi-interest extract layer in the user tower. Such a layer uses a neural network to dynamically determine the number of user embeddings and learns a mask vector to mask the undesired embedding representations. As a result, more active users would have more interest embeddings.
Moreover, we design a Prototypical Contrastive Learning module, where the key idea is to: 
(i) cluster on item embeddings; 
(ii) use cluster centroids to represent high-level embeddings of similar items, which are also termed as prototypes;
(iii) maximize inter-cluster distance and minimize intra-cluster distance in the embedding space with contrastive learning.
The Prototypical Contrastive Learning module can learn the high-level semantic structure of items by clustering, and encourages item embeddings to be closer within the same prototype and farther from different prototypes.

\begin{figure*}
\centering
\includegraphics[width=15.5cm]{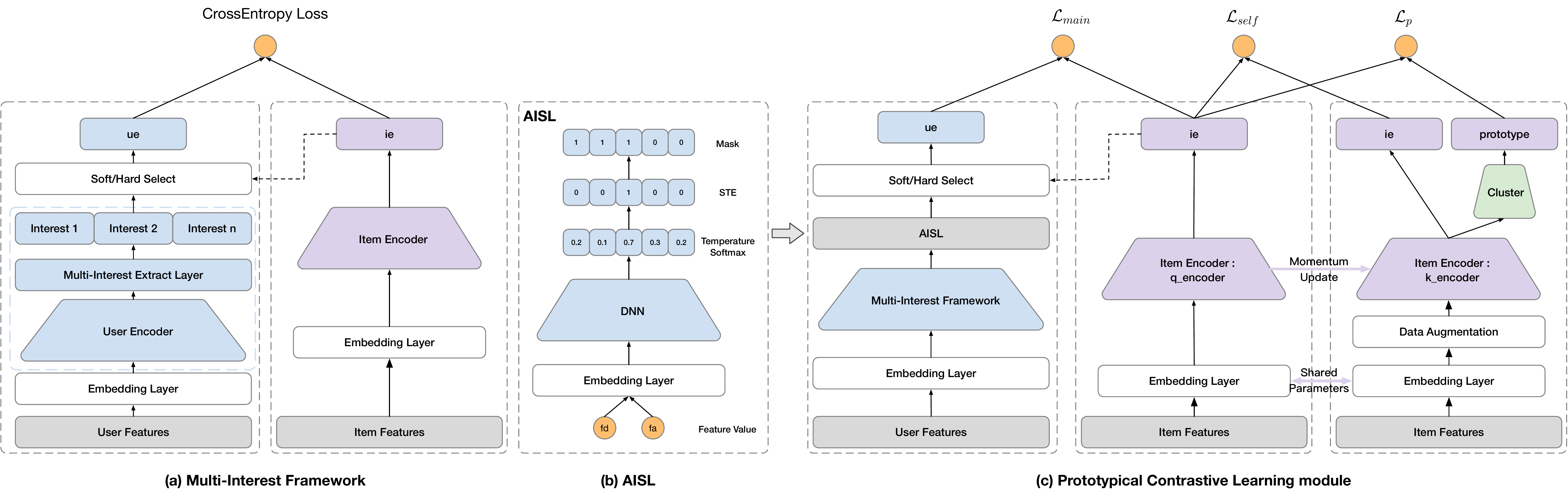}
\caption{Illustration of adaptive interest selection layer and prototypical contrastive learning module.}
\label{fig:picture001}
\end{figure*}

\section{PROPOSED MODEL}
\subsection{Problem Formulation}
Assume we have a set of users $u \in \mathcal{U}$ and a set of items $i \in \mathcal{I}$.
User $u$ has a sequence of historical behaviors in the form of interacted items $(i_1^u,i_2^u,\cdots,i_n^u)$, sorted by the time of the interaction. 
The DCG model involves learning a user encoder $ue=\mathcal{F}(x_u) \in \mathbb{R}^{K\times d}$, and an item encoder $ie=\mathcal{G}(x_i) \in \mathbb{R}^d$, where $x_u, x_i$ are the corresponding features, $K$ is the maximum number of user embeddings, and $d$ is the dimension of the embeddings. It then takes $ie$ and builds a KNN indexing service, e.g. Faiss \cite{JDH17faiss}. As a result, we can generate item candidates for user $u$ by finding the top-$N$ $ie$s closest to $ue$.
\begin{align}
    % \hat{y} &= \text{sigmoid}(s({ue}, {ie})) \label{eq:loss-y}\\
    \mathcal{L}_{main} &= - \frac{1}{B} \sum_{j\in[B]} [y_j \text{log} (\hat{y}_j) + (1-y_j) log(1-\hat{y}_j)] \label{eq:loss-ce}
\end{align}

In the end, the label is termed as $y$ and the predicted score
%for interaction between a user $u$ and an item $i$ 
is termed as $\hat{y} = \text{sigmoid}(s({ue}, {ie}))$, where $s$ is similarity measure function, e.g. cosine similarity. %, and inner product.
%which is shown in Eq \ref{eq:loss-y}.
For a batch of $B$ samples, we use cross entropy loss as is shown in Eq \ref{eq:loss-ce}.

\subsection{Adaptive Interest Selection Layer}
The multi-interest framework uses a multi-interest extract layer to extract multiple user embeddings as is shown in Figure \ref{fig:picture001}-a. 
Our proposed AISL is applied after the multi-interest extract layer as is shown in Figure \ref{fig:picture001}-b and is designed to adaptively learn the number of user embeddings.

%\subsubsection{Input and Output}
The actual number of user embeddings $K_u$ that AISL learns for an arbitrary user is highly relevant to the activeness of users. To allow AISL to have such knowledge, we take user demographic $f_b$ (e.g., age, gender) and activity features $f_a$ (e.g., the number of interactions of $u$) as the input of AISL. 
%\textbf{Output:} 
The output of AISL is a multi-hot mask vector $s_u \in \mathbb{R}^K$ for each user. 

%\subsubsection{Architecture}
Specifically, a multilayer perceptron of $L$ layers, is firstly applied to learn the user's activeness in latent space as is shown in Eq. \ref{eq:aisl-hl} where $h_0=f_b \odot f_a$. For notion, $\odot$ stands for concatenation of features and $\sigma$ is the activation function.
Then we apply $K$-temperature softmax layer on $h_L\in \mathbb{R}^K$ as is shown in Eq. \ref{eq:aisl-p} where $p_u \in \mathbb{R}^K$ denotes the probability of selecting different number of user embeddings $K_u$ from $1$ to $K$ for user $u$.
The $k$-th element of $p_u$ is termed as $p_{u}^k$. 
Thus, the multi-hot mask vector $s_u$ used for the user's interest selection can be obtained by Eq. \ref{eq:aisl-s}.
\begin{align}
    h_l &=\sigma(W_l^T h_{l-1} + b_l), \quad l \in [1,L] \label{eq:aisl-hl} \\ 
    p_{u}^k &= \text{exp}(h_{L}^k)/\sum_j \text{exp}(h_{L}^j) \label{eq:aisl-p} \\
    s_u &= {\text{multi\_hot}}(\arg\max_{k}(p_{u}^k)) \label{eq:aisl-s} 
\end{align}

A problem in this method is that the learning process is non-differentiable due to $\text{argmax}$ operation in Eq. \ref{eq:aisl-s}, which means we can't optimize the parameters by stochastic gradient descent (SGD) directly.
To solve this problem, we relax $s_u$ to a continuous space by temperature softmax and the $k$-th element of $s_u$, which is termed as $s_{u}^k$ can be calculated by Eq \ref{eq:aisl-suk}.
\begin{align}
    s_{u}^k \approx \hat{{s}_{u}^k} = (\text{exp}(h_{L}^k)/T)/ \sum_j (\text{exp}(h_{L}^j) / T) \label{eq:aisl-suk}
\end{align}
where $k \in [1, K]$ and $T \in \mathbb{R}^{+}$ is the temperature hyper-parameter.
However, this will introduce a gap between training using $\hat{s_u}$, and inference using $s_u$. To solve this inconsistency problem between online and offline, 
we rewrite $s_u$ as 
\begin{align}
    s_u = \hat{s}_u + \text{stop\_gradient}(s_u - \hat{s}_u)
\end{align}
$\text{stop\_gradient}$ operation only takes effect in the training stage and address the gap issue, which is inspired by the idea of previous work Straight-Through Estimator (STE) \cite{YoshuaBengio2013EstimatingOP}.

\subsection{Prototypical Contrastive Learning Module}
To address the class collision problem, we design the prototypical contrastive learning module as is shown in Figure \ref{fig:picture001}-c.

\subsubsection{Contrastive Learning Task}
% 解释数据增强
Data augmentation is an essential part in contrastive learning task and should be designed such that the augmented version from the same input can still be recognized by the models.
For this purpose, we choose a random mask data augmentation method to generate another representation version of the same item.
Specifically a mask layer is applied after the standard embedding layer's output $e_x$. 
The output of the mask layer $m_x \in \mathbb{R}^D$ only contains elements of ``0'' and ``1'', and is constrained to have a fixed number of ``1''s in possibly different positions for different $e_x$.
Thus, the augmented embedding $e_x'$ can be calculated by $e_x' = m_x e_x $.

% item Self-Supervised Learning(SSL)
We then design a contrastive learning algorithms to help to train the item embeddings.
The idea is two folds: first, we apply data augmentation transformation mentioned above for the items; 
and construct contrastive task between outputs of the query encoder and the key encoder.
The query encoder takes origin item feature embeddings $e_x$ as inputs, while the key encoder takes the augmented version $e_x'$ as inputs.
In DCG models, we treat the item tower as the query encoder and build the key encoder with the same neural network structure, which is driven by a momentum update with the query encoder and enables a large and consistent item representation learning as is shown in Figure \ref{fig:picture001}-c.
\begin{align}
    \text{query}_x &= \mathcal{G}(e_x) = \text{q\_encoder}(e_x) \label{eq:ssl-query} \\
    \hat{\text{key}_x} &= \mathcal{G}'(e_x') = \text{k\_encoder}(e_x'), \quad x \in {[0,J]} \label{eq:ssl-key}\\
    \text{key}_x &= \text{stop\_gradient}(\hat{\text{key}_x}) \\
    \mathcal{G}' &\gets \alpha \mathcal{G}' + (1-\alpha)\mathcal{G} \quad \alpha \in (0,1) \label{eq:ssl-m}
\end{align}

The item tower encoder is termed as $\text{q\_encoder}$ with parameters $\mathcal{G}$, which encodes original features to item embeddings as $\text{query}$ in Eq. \ref{eq:ssl-query}. 
The corresponding key encoder is termed as $\text{k\_encoder}$ with parameters $\mathcal{G}'$, which encodes the data augmented feature version to item embeddings as ${\hat{\text{key}}}$ in Eq. \ref{eq:ssl-key} and is updated in a momentum way according to Eq. \ref{eq:ssl-m}.
We use a hyper-parameter $\alpha$, which is close to 1, to control the updating speed.
For each item interacted by the user, we randomly selected ${J}$ size items to form the negative pairs for the contrastive task. Thus, we get one pair of positive sample, the interacted item and it's data augmented version, and ${J}$ pairs of negative samples, the interacted item and the data augmented results of ${J}$ randomly selected items.

The contrastive learning task encourages $\text{query}_x$ to be similar to $\text{key}_x$ for the same item candidate, but to keep distance from $\text{key}_x$ of other items. The self-supervised learning loss function for a batch of $B$ samples can be expressed as 
\begin{align}
    \mathcal{L}_{self}(\{e_x\};\mathcal{G},\mathcal{G}') = - \frac{1}{B} \sum_{x\in[B]} \text{log} \frac{\text{exp}(s(e_x, e_x')/T)}{\sum_{j\in[B]}\text{exp}(s(e_x, e_j')/T)} \label{eq:ssl-loss}
\end{align}
where $T$ is temperature hyper-parameter.

\subsubsection{Prototype Augmented Module}
% 引入prototype，增加item维度 global辅助loss
A prototype $c_x \in \mathcal{C}$ is defined as the cluster centroid for a group of semantically similar items. The clusters result from the output of the key encoder $\mathcal{G}'(e_x')$ by K-means \cite{JamesBMacQueen1967SomeMF}, and the number of the clusters is fixed to $|\mathcal{C}|$.
%which is the output of $\mathcal{G}'(e_i')$ of all the candidate items. 
Then, we construct a prototype-based contrastive loss as in Eq. \ref{eq:proto-loss}. It enforces the embedding of an item $e_x$ to be more closer to its corresponding prototype $c_x$ compared to other prototypes, which are $r$ randomly chosen prototypes from clusters $e_x$ does not belong to.

For a prototype $c$ with $N_c$ items assigned, we use $\tau_c$ to replace the temperature parameter in softmax as is shown in Eq. \ref{eq:proto-tao}. This has been proved to yield prototypes with similar amount of items and lower variance of item embeddings \cite{JunnanLi2020PrototypicalCL}.
\begin{align}
    \mathcal{L}_{p} &= - \frac{1}{B} \sum_{x\in[B]} \text{log} \frac{\text{exp}(s(e_x, c_x)/\tau_{c_x})}{\text{exp}(s(e_x, c_x)/\tau_{c_{x}}) + \sum_{j\in[r]} \text{exp}(s(e_x, c_j)/\tau_{c_j}) } \label{eq:proto-loss} \\
    \tau_c &= \frac{\sum_{x \in [N_c]}{||e_x'-c_x||}_2}{ N_c \text{log}(N_c + 1) } \quad c \in \mathcal{C} \label{eq:proto-tao}
\end{align}

% \subsubsection{End-to-end Framework}
We adjointly train prototypes with user or item embeddings and optimize model parameters in an end-to-end framework. The total loss function is
\begin{align}
    \mathcal{L} = \mathcal{L}_{main} + \alpha \mathcal{L}_{self} + \beta \mathcal{L}_p
\end{align}
and $\alpha,\beta$ are hyper-parameters.

\begin{table*}[ht]
% \begin{table*}[H]  
	\centering
% 	\caption{Recommendation performance on public datasets. The best results are highlighted with bold fold. All the numbers in the table are percentage numbers with ``\%'' omitted. AINPU stands for average interest number per user.} 
	\caption{The best results are highlighted with bold fold. All the numbers are percentage numbers with ``\%'' omitted.} 
	\label{table:methodcompare}
	\begin{tabular}{c|cccccc|cccccc}
		\toprule[1pt]
		\multirow{1}*{} & \multicolumn{6}{c}{\textbf{Amazon Books}} &  \multicolumn{6}{c}{\textbf{Kindle Store}} \\ 
		& \multicolumn{3}{c}{Metrics@10} & \multicolumn{3}{c}{Metrics@20} & \multicolumn{3}{c}{Metrics@10} & \multicolumn{3}{c}{Metrics@20} \\

        \hline
		& AINPU & Hit Rate & NDCG & AINPU & Hit Rate & NDCG & AINPU & Hit Rate & NDCG & AINPU & Hit Rate & NDCG\\
		\hline 
		\textbf{MostPopular} & - & 10.59 & 8.53 & - & 15.39 & 10.93  & - & 12.42 & 10.06 & -  & 18.30 & 13.61\\

		\textbf{YouTube DNN} & 1 & 22.67 & 14.67 & 1 & 38.28 & 15.07 & 1 & 24.75 & 14.11 & 1  & 39.10 & 26.24\\
		\textbf{MIND}  & 5 & 30.05 & 22.80 & 5 & 42.93 & 24.36 & 4 & 35.19 & 20.94 & 4  & 45.84 & 29.39\\
		\textbf{SASRec}  & 5 & 31.86 & 22.99 & 5 & 42.61 & 24.23 & 4 & 35.54 & 22.09 & 4  & 45.06 & 28.50\\

        \hline
		\textbf{Our approach}  & {\bf 4.1}  & {\bf 34.75} & {\bf 25.54} & {\bf 4.1} & {\bf 45.10} & {\bf 27.52}  & {\bf 3.2} & {\bf 37.47} & {\bf 25.07} & {\bf 3.2}  & {\bf 47.36} & {\bf 32.27}\\
		\bottomrule[1pt]
	\end{tabular}
\end{table*}
% 参考：https://blog.csdn.net/brave_stone/article/details/88931187
% Average Interest Number Per User AINPU

\section{EXPERIMENTS}

\subsection{Experimental Setup}

\noindent\textbf{Dataset.} We conduct experiments on an \textbf{Amazon Dataset} $\footnote{https://jmcauley.ucsd.edu/data/amazon/}$. 
This dataset contains product reviews and metadata from Amazon and each sub-category dataset consists of meta information of items as well as users' historical behaviors on items. We use two sub-categories in this dataset to evaluate our method including Books and Kindle Store. We follow the data process method used in DIEN\cite{GuoruiZhou2019DeepIE}. The details of dataset are summarized in Table \ref{table:dataset}.

\noindent\textbf{Competitors.} We compare our proposed models with the state-of-the-art models. \textbf{MostPopular} is a classic strategy that recommends most popular items to users in recommender systems. 
\textbf{YouTube DNN} \cite{YouTubeDNN} represents the single user embedding models in DCG.
\textbf{MIND} \cite{MIND} and \textbf{SASRec} \cite{WangChengKang2018SelfAttentiveSR} are recently proposed multi-interest methods based on capsule networks \cite{SaraSabour2017DynamicRB} and multi-head self-attention \cite{AshishVaswani2017AttentionIA}, respectively.
\begin{table}[H]
    \caption{Statistics of datasets.}
    % \vspace{10pt}
    \centering
    \begin{tabular}{c|c|c|c}
        \hline
        \hline
        \textbf{Dataset} & \# users & \# items & \# interactions \\
        \hline
        Kindle Store & 61,235& 62,528 & 123,108 \\
        Amazon Books & 459,133 & 313,966 & 8,898,041 \\
        \hline       
    \end{tabular}
    \label{table:dataset}
\end{table}

\noindent\textbf{Parameters.} For a fair comparison, all methods are implemented in Tensorflow and use Adam optimizer with $\alpha=0.01,\beta_1=0.9,\beta_2=0.999$, a mini-batch size of 256 and a fixed learning rate as 0.001. We set the dimension of user and item embeddings $d$ as 32, the max number of user embeddings $K$ as 5 for Amazon Books and 4 for Kindle Store respectively according to the number of interactions in the dataset. 
%And we set cluster number $\mathcal{C}=1000$, which is sufficient for the scale of the dataset. 
The cluster number $|\mathcal{C}|$ is set to 1000, which is sufficient for the scale of the dataset. Each item samples $r=100$ negative prototypes. The parameters of the competitors are tuned according to values suggested in the original papers.

\noindent\textbf{Metrics.} For each user in the test set, we sample 100 items randomly from all items as negative samples, which the user has not interacted with and rank them together with the observed positive samples.
% We use two commonly used metrics, Hit rate (HR)@N and Normalized Discounted Cumulative Gain (NDCG)@N, to evaluate the performance.
% There are two commonly used metrics to evaluate the performance of DCG models.
Hit rate (HR)@N measures the percentage that retrieved items contain at least one positive item interacted by the user
%, which has been widely used in the literature 
 \cite{GeorgeKarypis2001EvaluationOI}. 
Normalized Discounted Cumulative Gain (NDCG) is a measure of ranking quality that is often used to measure effectiveness of information retrieval systems. It returns a high value if positive items are ranked high by predicted scores\cite{KalervoJrvelin2000IREM}.
We also record the average number of user embeddings (AINPU). With similar HR or NDCG results, a lower AINPU indicates more efficiency.

\subsection{Experimental Results}
We conduct extensive experiments on \textbf{Amazon Dataset}, and report the results in Table \ref{table:methodcompare}. 
Our method outperforms all the state-of-the-art models by a wide margin on all the evaluation metrics. 
% 多向量效果更好
In competitors, SASRec obtains the best performance and it gains 9.19\% improvements in HR@10, 8.32\% in NDCG@10 over YouTube DNN which only outputs single user embedding. MIND gets comparable results with SASRec. It shows that multi-interest models have much better abilities to capture user's diverse interests than single-interest models. 
% 我们的方法更好，并且AINPU更低，说明固定数量的用户多向量有多余的部分，学习不够准确，召回了unrelated items
Compared to the best competitor SASRec, our method gains 2.89\% improvements in HR@10, 3.77\% in NDCG@20 and the AINPU is relatively reduced by 18\% from 5 to 4.1 in Amazon Books and by 20\% from 4 to 3.2 in Kindle Store.
Our method achieves a lower AINPU but higher HR@10 and NDCG@20, which means AISL addresses the aforementioned issues in multi-interest framework that the number of user embeddings should be personalized. 
Especially, AISL reduces the redundant user embeddings, which hurts the candidate generation performance for non-active users. Essentially, a lower AINPU helps to improve the efficiency of the system. 
\begin{figure}[H]
	\centering
	\includegraphics[width=9cm]{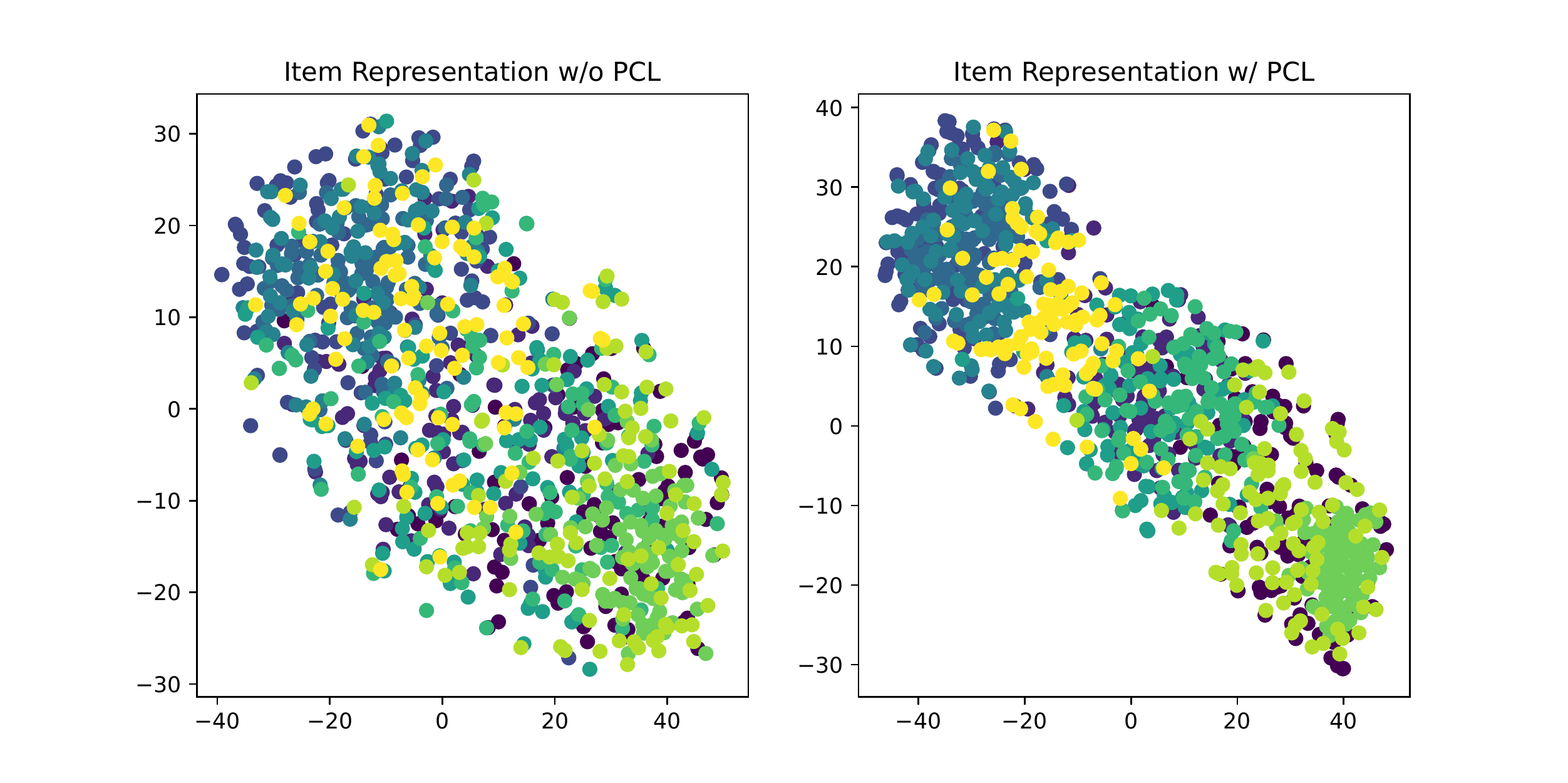}
% 	\includegraphics[width=9cm]{representation.pdf}
% 	\includegraphics[width=8cm]{representation.pdf}
% 	\includegraphics[width=7.5cm]{representation.pdf}
% 	\caption{Randomly select 10 categories with 100 samples per category and draw T-SNE visualization of the item embeddings. Colors represent categories.}
	\caption{T-SNE of item embeddings: randomly select 10 categories with 100 samples each. Colors represent categories.}
	\label{fig:representation}
\end{figure}

We also study the influence of the Prototypical Contrastive Learning Module as illustrated in Figure \ref{fig:representation}.
It is obvious that items within the same category have much closer embedding representations with the proposed module applied. It benefits from the contrastive learning task and the proposed prototype-based module which can be regarded as a special case of data augmentation. Our method effectively addresses the class collision problem.

\section{CONCLUSION}
In this paper, we designed an Adaptive Interest Selection Layer to adaptively learn the number of user embeddings and reduce redundant representations, which is more effective and efficient in a multi-interest candidate generation framework. 
%makes the number of user interests personalized
We also proposed Prototypical Contrastive Learning Module, which introduces prototypes by clustering item embeddings to alleviate the class collision problem.
We evaluated the proposed method on public datasets compared with the state-of-the-art competitors, and our proposed method showed significant performance improvements under three evaluation metrics.

\bibliographystyle{ACM-Reference-Format}
\balance
\bibliography{pcs.bib}

\end{document}